\documentclass[a4paper]{jpconf}
\usepackage{graphicx}
\begin{document}
\title{Intergalactic $\gamma$-ray propagation: \\ basic ideas, processes, and constraints}

\author{Timur Dzhatdoev$^{1,2}$, Emil Khalikov$^{1}$, Egor Podlesnyi$^{3,1}$ \\ and Anastasia Telegina$^{3}$}
\address{$^{1}$ Skobeltsyn Institute of Nuclear Physics, Moscow State University, Moscow, 119991 Russia}
\address{$^{2}$ Institute for Cosmic Ray Research, University of Tokyo, 5-1-5 Kashiwanoha, Kashiwa, Japan}
\address{$^{3}$ Faculty of Physics, Moscow State University, Moscow, 119991 Russia}
\ead{timur1606@gmail.com, nanti93@mail.ru, podlesnyi.ei14@physics.msu.ru, nastias.skateme@gmail.com}

\begin{abstract}
We review extragalactic $\gamma$-ray propagation models with emphasis on the electromagnetic (EM) cascade process in the magnetized expanding Universe. We consider cascades initiated by primary protons of ultra-high energy accelerated by blazars and show that the observable spectrum is similar to the universal spectrum of a purely EM cascade. We also present a detailed calculation of the observable angular distribution for the case of EM cascades developing from relatively nearby ($<$20 Mpc) sources. Finally, we calculate the point-like source differential sensitivity of a novel liquid Argon time projection chamber $\gamma$-ray telescope and show that its sensitivity is several times better than the Fermi LAT sensitivity in the 100 MeV -- 100 GeV energy range.
\end{abstract}

\section{Introduction}

Observable $\gamma$-ray spectral, angular, and temporal distributions for extragalactic sources may be significantly transformed during intergalactic propagation. The most basic elementary processes involved are: \\
1) electron-positron pair production (PP) $\gamma\gamma \rightarrow e^{+}e^{-}$ on extragalactic background light (EBL) \cite{nik62,gou67} and cosmic microwave background (CMB) photons \cite{jel66} for $\gamma$-rays; \\
2) inverse Compton (IC) scattering $e\gamma \rightarrow e^{'}\gamma{'}$ for electrons and positrons\footnote{hereafter simply ``electrons''} that occurs mainly on the CMB (see \cite{blu70} and references therein); \\
3) Bethe-Heitler pair production (BHPP)  and pion photoproduction (PIP) for protons and nuclei\footnote{below we do not consider primary nuclei, thus the photodissociation process is not relevant}. For ultrahigh energy (UHE, $E>1/(1+z)$ EeV, $z$ is redshift) BHPP is usually dominated by interactions on the CMB. For $1/(1+z)$ EeV $<E<50/(1+z)$ EeV PIP is subdominant; at higher energies it is the main proton energy loss process, occuring mainly on the CMB (see \cite{ber06} and references therein); \\
4) adiabatic losses (AL). \\
For especially high values of center-of-mass energy, additional processes also set in, such as muon pair production $\gamma\gamma \rightarrow \mu^{+}\mu^{-}$  \cite{li07} and double pair production $\gamma\gamma \rightarrow e^{+}e^{-}e^{+}e^{-}$ \cite{bro73} for $\gamma$-rays. Additionally, there is a quest for New Physics processes, such as $\gamma$-axion-like
particle (ALP) mixing \cite{raf88}--\cite{ana17} or Lorentz invariance violation (LIV) \cite{kif99}--\cite{abd18}. 

Models involving production and propagation of $\gamma$-rays in the intergalactic volume may be classified as belonging to one of the following three types \cite{dzh17a,dzh17b}: \\
1) the absorption-only model, which accounts for only the PP and AL processes; \\
2) cascade models, which also include the IC process, and BHPP and PIP for primary protons; \\
3) exotic models. \\
Following the lines of argument presented in \cite{dzh17b}, in this paper we concentrate on the cascade models. To recall, the physics behind the absorption-only model is already very well-known. Many exotic models are already ruled out \cite{aha00} \cite{rub17}, and there is growing evidence against some other options such as $\gamma$-ALP mixing  \cite{aj16,mal18}. On the other hand, all existing possible deviations from the absorption-only model, such as \cite{hor12,kor18a} (see also \cite{fur15,che15}), could be qualitatively accomodated in the framework of the cascade models \cite{dzh17}. A caveat is that these anomalies are still not very well established \cite{bit15,dom15} (see also \cite{fer18}).

The present paper is organised as follows. In section 2 we consider two specific cases where intergalactic electromagnetic (EM) cascades are important, namely: secondary $\gamma$-rays from 1) ultra-high energy cosmic rays emitted by blazars (subsection 2.1) and 2) relatively nearby sources such as galaxy clusters with the source-observer distance less than 20 Mpc (subsection 2.2). A number of other case studies could be found in \cite{dzh17,dzh17a,dzh18}. In section 3 we discuss a new experimental technique called on to measure the extragalactic magnetic field (EGMF), namely, the liquid argon time projection chamber (TPC) approach. Finally, we conclude in section 4. All figures presented below were produced with the ROOT software \cite{bru97}.

\section{Specific cases}

\subsection{Intergalactic hadronic cascade model}

$\gamma$-rays from EM cascades initiated by primary protons accelerated in blazars may contribute to the observable emission (see e.g. \cite{ury98,ess11}). These protons may be strongly deflected on large scale structure (LSS) filament magnetic fields; below we show that this effect impacts the observable $\gamma$-ray spectrum (see also \cite{dzh17}). A simplified scheme of the corresponding geometry is shown in figure 1. A source ($S$) emits UHE protons that first propagate through a void (underdense region of space) with the diameter $L_{V}$ and then are deflected on a filament (denoted by twin dashed lines). The proton deflection angle is denoted as $\delta$, $\theta_{obs}$ is the angle between the direction from the source to the observer ($O$) and the direction of an incoming observable $\gamma$-ray. $L$ is the distance between the source and the observer, $L_{int} = L_{V} + L_{f}$ is the characteristic proton interaction length.

\begin{figure}[h]
\centering
\includegraphics[width=15cm]{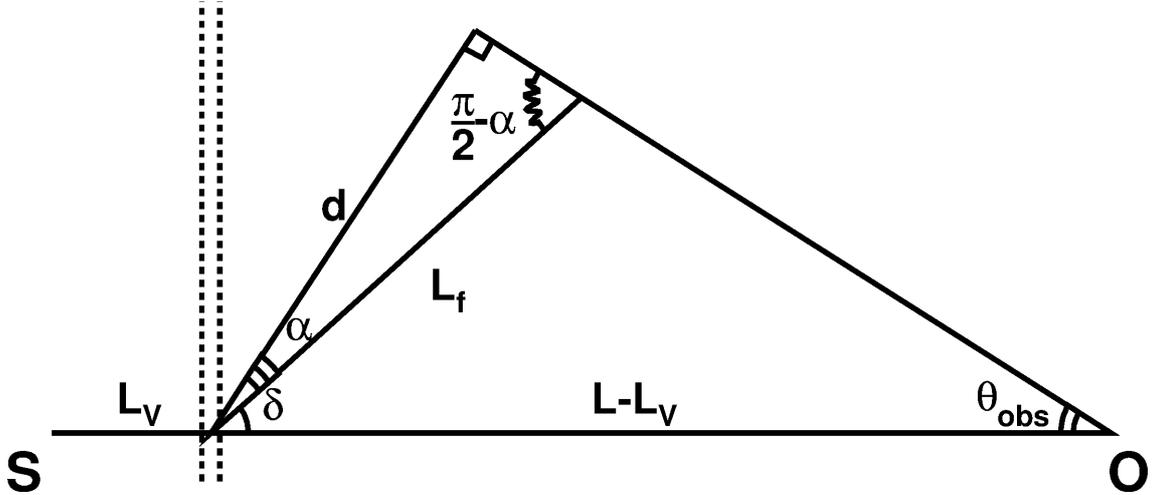}
\caption{A scheme of geometry for the hadronic cascade model (not to scale).}
\label{fig1}
\end{figure}

Let's extrapolate the observable $\gamma$-ray path until we have a triangle with a right angle. The shortest side of that triangle can be represented by the following equation:
\begin{equation}
d = \sin(\theta_{obs}) \cdot (L - L_{V}) = \sin(\frac{\pi}{2} - \alpha) \cdot L_{f}
\end{equation}
At the same time
\begin{equation}
\delta + \alpha = \frac{\pi}{2} - \theta_{obs} \rightarrow \frac{\pi}{2} - \alpha = \delta + \theta_{obs}
\end{equation}
Assuming that $\theta_{obs} \ll \frac{\pi}{2}$ and $\delta \ll \frac{\pi}{2}$ we obtain the following:
\begin{equation}
\theta_{obs} \cdot (L - L_{V}) \approx (\delta + \theta_{obs}) \cdot L_{f}
\end{equation}
\begin{equation}
\theta_{obs} \cdot (L - L_{f} - L_{V}) \approx \delta \cdot L_{f}
\end{equation}
And finally
\begin{equation}
\theta_{obs} \approx \frac{\delta \cdot L_{f}}{L - L_{f} - L_{V}}= \delta \frac{L_{int}-L_{V}}{L - L_{int}}.
\end{equation}
Now let's estimate $\theta_{obs}$ assuming $L = 750$ Mpc and $L_{V} = 50$ Mpc. Let's assume $L_{int}$ to be one of the following set of values: $100, 200$ and $500$ Mpc. The proton deflection angle $\delta$ may be estimated as follows \cite{har16}:
\begin{equation}
\delta \approx 1^{\circ} \frac{B}{nG} \frac{40 EeV}{E/Z} \frac{\sqrt{L_{B}l_{c}}}{Mpc},
\end{equation}
where $B$ is the magnetic field strength, $E$ and $Z$ are the energy and charge of the primary particles (protons in our case), $L_{B}$ is the thickness of the filament and $l_{c}$ is the coherence length of that magnetic field. For $B = 1$ nG, $E = 40$ EeV, $L_{B}$ = 1 Mpc, $l_{c} = 1$ Mpc, we get $\delta$= 1 $^{\circ}$ and $\theta_{obs}= 0.077^{\circ}, 0.27^{\circ}, 1.8^{\circ}$ for the three values of $L_{int}$. The typical extension of an imaging atmospheric Cherenkov telescope (IACT) point spread function (PSF) is about 0.1$^{\circ}$, thus a part of observable $\gamma$-ray flux in our case will not fit into the point-like image of the source. In a more realistic case of lower proton energies, lesser values of $L_{V}$, and many filaments on the line of sight, this leads to the conclusion that most of observable $\gamma$-rays produced farther than 100 Mpc from the source would not contribute to the observable spectrum of the source. This, in turn, leads to an effective cutoff at the highest observable energies, as pointed out in \cite{dzh17}. In effect, the observable spectrum is almost the same as in the framework of the electromagnetic cascade model in the universal regime \cite{ber16}.

\subsection{EM cascades from nearby sources}

Another interesting, rarely discussed case is that of relatively nearby sources ($L<$20 Mpc). Here we consider an example of the Virgo cluster located at $L$=16.8 Mpc from the Earth. Very recently, \cite{bla18} considered this source in context of very heavy dark matter searches. In particular, the authors of \cite{bla18} estimated the secondary electron deflection angle (see their equation (3.7)). We note that this equation was obtained following e.g. \cite{ner09}. The latter work assumed blazars as sources of primary $\gamma$-rays. Blazars typically have $z>$0.03, therefore the energy of last-generation cascade electrons is modest ($\sim$10 TeV or less), making the Thomson approximation used by \cite{ner09} fully applicable in this case. For nearby sources, however, such electrons may have the energy well in excess of 10 TeV, where the Thomson approximation is not valid. Figure 2 shows the electron mean free path (black curve), as well as the characteristic electron energy loss length $L_{E-e}= c\cdot E/(dE/dt)$ (blue curve). These quantities were computed using the approximation of \cite{kha14}. For comparison, red line denotes $L_{E-e}$ in the Thomson approximation. We note that for electron energy in excess of 50 TeV the Thomson approximation fails and thus a new calculation of observable spectral and angular distributions is required.

Using the publicly-available code of \cite{fit17} and assuming the EGMF strength 1 nG on coherence length 1 Mpc, we have performed a detailed calculation of the observable angular distribution. Such EGMF parameters were promtped by the fact that we do not expect a large void with low magnetic field strength in the local Universe. We also assumed the EBL model of \cite{gil12} which is consistent with current constraints such as \cite{kor18b}. Figure 3 shows the observable angular distribution for three ranges of primary energy: 20--36 TeV (black histogram), 97--158 TeV (red histogram), and 1--3 PeV (blue histogram). In the first case the primary $\gamma$-ray mean free path $L_{\gamma}>L$; the second case corresponds to the one-generation regime, and the last case --- to the universal regime (see \cite{dzh17} for detailed discussion of the latter two regimes). We note that the typical total electron deflection angle is well in excess of $\pi$; therefore, EM cascades develop an almost isotropic cloud around the source --- the so-called pair halo (PH) \cite{aha94}.

\begin{figure}[h]
\begin{minipage}{18.5pc}
\includegraphics[width=18.5pc]{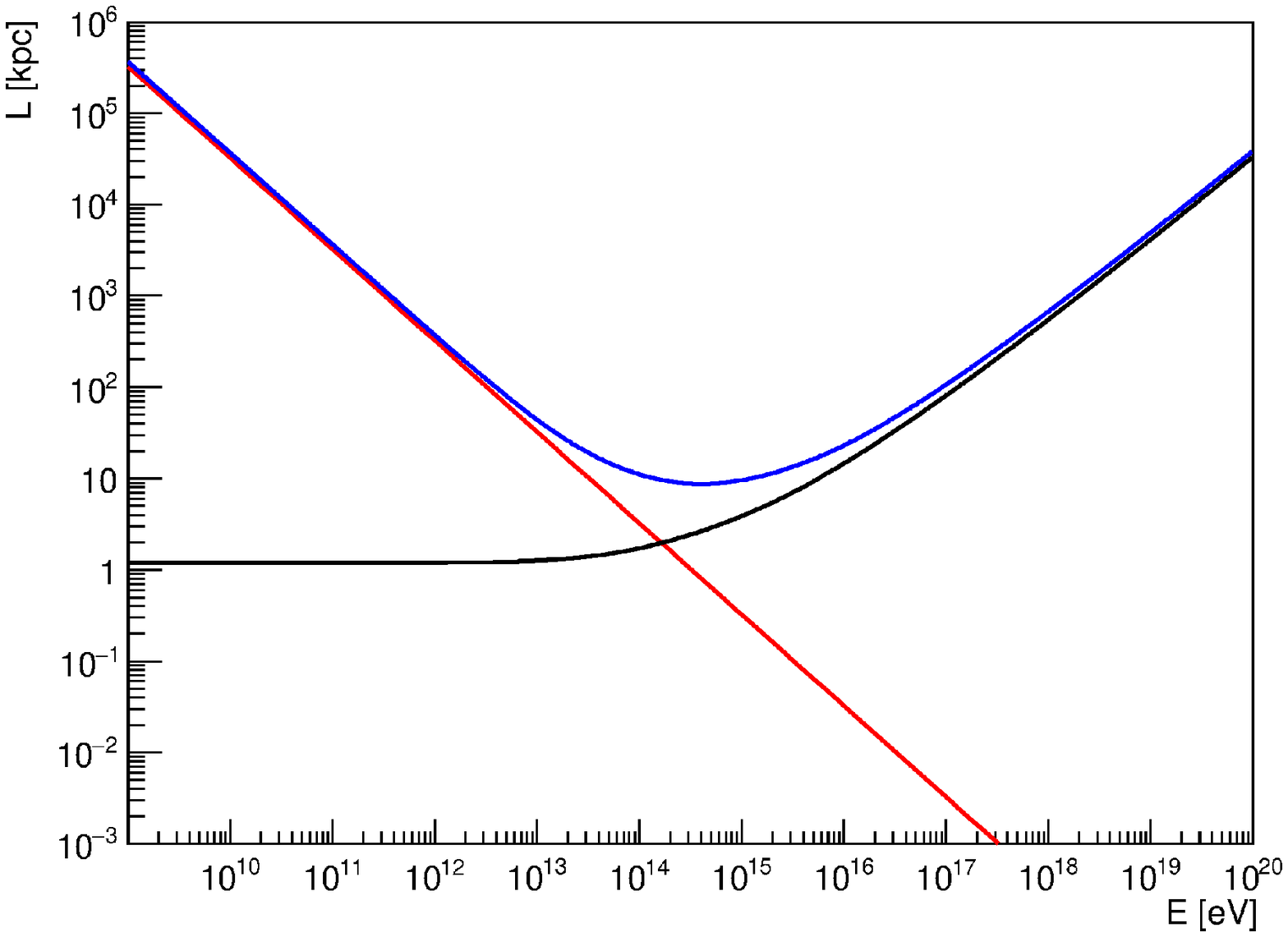}
\caption{Characteristic lengths for electron in the intergalactic medium at $z$=0 vs. energy.}
\label{fig2}
\end{minipage} \hspace{1pc}
\begin{minipage}{18.5pc}
\includegraphics[width=18.5pc]{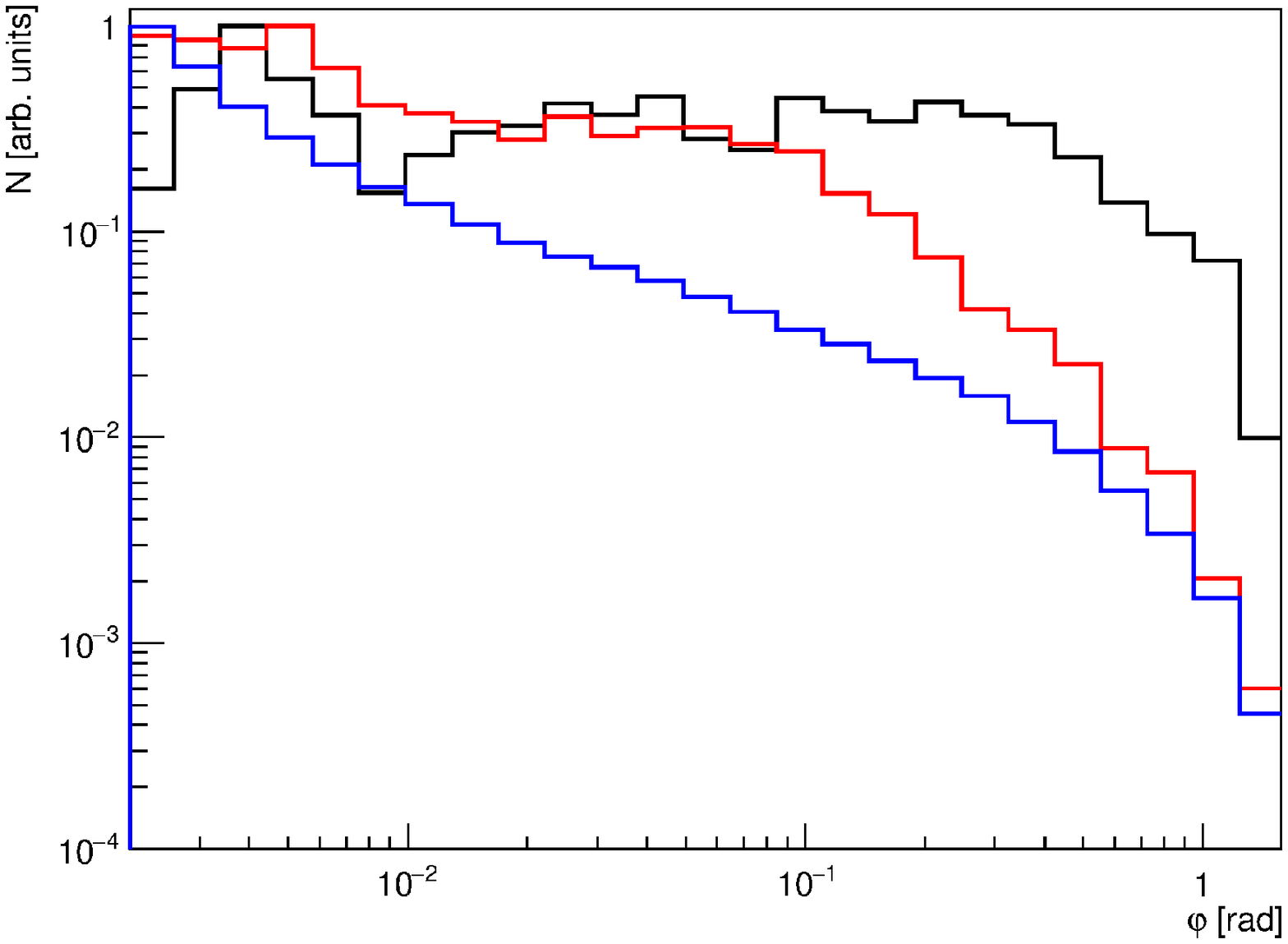}
\caption{Observable angular distributions for various primary energy ranges.}
\label{fig3}
\end{minipage}
\end{figure}

\section{Point-like source differential sensitivity of liquid Argon TPC $\gamma$-ray telescope}

The main uncertainty of extragalactic $\gamma$-ray propagation models is the unknown strength $B$ and structure of the EGMF (for a review, see \cite{dur13,han17}). Possible values of $B$ in voids range from 1 nG to 0.01 fG. Additionally, the EGMF may be highly inhomogeneous (see, e.g. \cite{hac18}).

In \cite{dzh18} we argue that the most robust method to measure weak ($B<$10 fG) EGMF in voids is to detect magnetically broadened cascade (MBC) emission from blazars. If $B<$1 fG, the angular width of the MBC is typically comparable or smaller than the PSF width of currently operating $\gamma$-ray instruments. Therefore, a new experimental technique with better angular resolution is required in order to measure weak EGMF. In \cite{dzh18} we propose, for the first time, to utilize a novel technique of liquid argon time projection chamber (TPC) \cite{ber13} to constrain the EGMF parameters. Here we provide an estimate of point-like source differential sensitivity vs. energy for such a detector following \cite{ber13}.

We assume a 2.0 m $\times$ 2.0 m $\times$ 0.8 m sensitive volume with 100 mkm longitudinal sampling step, filled with liquid Argon, segmented into 8 layers, each 10 cm thick. The total sensitive mass of the detector is 4480 kg, its total thickness is 110 g/cm$^{2}$ corresponding to 5.7 radiation lengths and optical depth $\tau=$3.3 for 100 MeV $\gamma$-rays. Currently we are working to develop an energy reconstruction technique utilizing such a detector of medium thickness.

Figure 4 shows sensitivity for this $\gamma$-ray instrument assuming 1.7 years of continious observation of a source (thick red curve) or 10 years of continious observation (blue thick curve) for 5 $\sigma$ detection of the source. Additionally, a minimum of 10 signal events are required; we also require that the number of signal events exceeds the number of background events. Other sensitivity curves were taken from \cite{fun13} (see also \cite{cta13}) and denote the differential sensitivity of Fermi LAT and atmospheric Cherenkov telescopes (ACT) \cite{act11,ach13} (see Figure 5 (left) of \cite{dzh17a} for the meaning of these curves). We conclude that the proposed new $\gamma$-ray instrument has the sensitivity several times greater than the Fermi LAT one in the wide energy range 100 MeV-- 100 GeV and could provide a natural low-energy extension for IACT such as the CTA array \cite{act11,ach13} or LHAASO \cite{lha14}.

\begin{figure}[h]
\centering
\includegraphics[width=12cm]{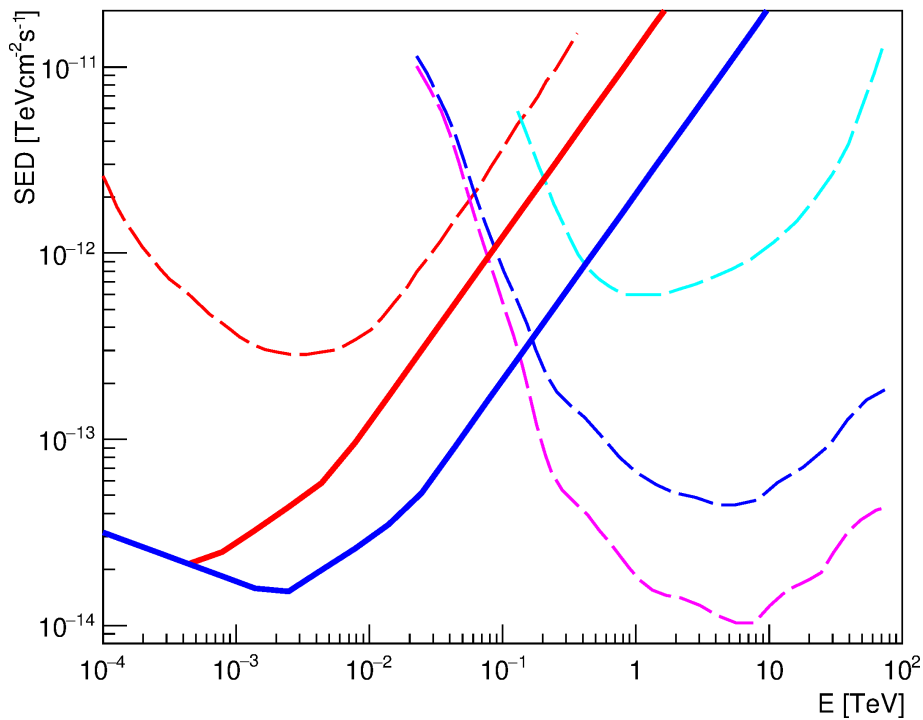}
\caption{Differential sensitivity for point-like sources for liquid argon TPC and other $\gamma$-ray instruments.}
\label{fig4}
\end{figure}

\section{Conclusions}

In this work we continued our review of extragalactic $\gamma$-ray propagation phenomenology concentrating on intergalactic EM cascades. We have shown that the typical observable spectrum in the framework of the intergalactic hadronic cascade model practically coincides with the universal spectrum of the electromagnetic cascade model. This is due to deflection of primary protons in LSS filaments, causing the broadening of the observable angular distribution. We note that this effect is relevant even for isotropic sources. We also considered EM cascades from relatively nearby ($L<$20 Mpc) sources such as galaxy clusters and calculated the observable angular distribution in this case. Finally, we have estimated the point-like source differential sensitivity for a novel liquid Argon TPC $\gamma$-ray telescope, finding out that it has the sensitivity several times better than Fermi LAT in the 100 MeV -- 100 GeV energy range.

\section*{Acknowledgements}

This work is supported by the Russian Science Foundation (RSF) (project No 18-72-00083).


\section*{References}
\begin{thebibliography}{99}
\bibitem{nik62}
Nikishov A I 1962 {\it Sov. Phys. JETP} {\bf 14} 393
\bibitem{gou67}
Gould R J and Shreder G 1967 {\it Phys. Rev.} {\bf 155} 1408
\bibitem{jel66}
Jelley J V 1966  {\it Phys. Rev. Lett.} {\bf 16} 479
\bibitem{blu70}
Blumenthal G 1970 {\it Phys. Rev. D} {\bf 1} 1596
\bibitem{ber06}
Berezinsky V, Gazizov A and Grigorieva S 2006 {\it Phys. Rev. D} {\bf 74} 043005
\bibitem{li07}
Li Z and Waxman E 2007 EeV neutrinos associated with UHECR sources  {\it Preprint} arXiv:0711.4969
\bibitem{bro73}
Brown R W et al. 1973 {\it Phys. Rev. D} {\bf 8} 3083
\bibitem{raf88}
Raffelt G and Stodolsky L 1988 {\it Phys. Rev. D} {\bf 37} 1237
\bibitem{kar17}
Kartavtsev A et al. 2017 {\it JCAP} {\bf 01} 024
\bibitem{mon17}
Montanino D et al. 2017 {\it Phys. Rev. Lett.} {\bf 119} 101101
\bibitem{ana17}
Anastassopoulos V et al. 2017 {\it Journal of Instrumentation} {\bf 12} P11019
\bibitem{col99}
Coleman S and Glashow S 1999 {\it Phys. Rev. D} {\bf 59} 11
\bibitem{kif99}
Kifune T 1999 {\it ApJ} {\bf 518} L21
\bibitem{tav16}
Tavecchio F and Bonnoli G 2016 {\it A\&A} {\bf 585} A25
\bibitem{abd18}
Abdalla H and Boettcher M 2018 {\it ApJ} {\bf 865} 159
\bibitem{dzh17a}
Dzhatdoev T 2017 {\it Proc. of the Moriond-2017 Very High Energy Phenomena in the Universe conf.} La Thuile
\bibitem{dzh17b}
Dzhatdoev T, Khalikov E and Kircheva A 2017 in {\it Proc. of Science, the 35th Int. Cosmic Ray Conf. 2017 (ICRC-2017)} 866 Busan
\bibitem{aha00}
Aharonian F A et al. 2000 {\it ApJ} {\bf 543} L39
\bibitem{rub17}
Rubtsov G, Satunin P and Sibiryakov S 2017 {\it JCAP} {\bf 1705} 049
\bibitem{aj16}
Ajello M et al. (Fermi-LAT Collaboration) 2016 {\it Phys. Rev. Lett.} {\bf 116} 161101
\bibitem{mal18}
Malyshev D et al. 2018 Improved limit on axion-like particles from gamma-ray data on Perseus cluster {\it Preprint} arXiv:1805.04388
\bibitem{hor12}
Horns D and Meyer M 2012 {\it JCAP} {\bf 1202} 033
\bibitem{kor18a}
Korochkin A, Rubtsov G and Troitsky S 2018 Distance-dependent hardenings in gamma-ray blazar spectra corrected for the absorption on the extragalactic background light {\it Preprint} arXiv:1810.03443
\bibitem{fur15}
Furniss A et al. 2015 {\it MNRAS} {\bf 446} 2267
\bibitem{che15}
Chen W et al. 2015 {\it Phys. Rev. Lett.} {\bf 115} 211103
\bibitem{dzh17}
Dzhatdoev T, Khalikov E, Kircheva A and Lyukshin A 2017 {\it A\&A} {\bf 603} A59
\bibitem{bit15}
Biteau J and Williams D 2015 {\it ApJ} {\bf 812} 60
\bibitem{dom15}
Dominguez A, Ajello M 2015 {\it ApJ} {\bf 813} L34
\bibitem{fer18}
Fermi-LAT Collaboration and Biteau J 2018 {\it ApJ Supplement Series} {\bf 237} 32
\bibitem{dzh18}
Dzhatdoev T et al. 2018 Intergalactic electromagnetic cascades in the magnetized Universe as a tool of astroparticle physics {\it Preprint} arXiv: 1808.06758
\bibitem{bru97}
Brun R and Rademakers F 1997 {\it Nucl. Instrum. Methods Phys. Res. Sec. A} {\bf 389} 81
\bibitem{ury98}
Uryson A 1998 {\it JETP} {\bf 86} 213
\bibitem{ess11}
Essey W et al. 2011 {\it ApJ} {\bf 731} 51
\bibitem{har16}
Harari D, Mollerach S and Roulet E 2016 {\it Phys. Rev. D} {\bf D93} 063002
\bibitem{ber16}
Berezinsky V and Kalashev O 2016 {\it Phys. Rev. D} {\bf 94} 023007
\bibitem{bla18}
Blanco C, Harding J P and Hooper D 2018 {\it JCAP} {\bf 1804} 060
\bibitem{ner09}
Neronov A and Semikoz D 2009 {\it Phys. Rev. D} {\bf 80} 123012
\bibitem {kha14}
Khangulyan D, Aharonian F A and Kelner S R. 2014 {\it ApJ} {\bf 783} 100
\bibitem{fit17}
Fitoussi T et al. 2017 {\it MNRAS} {\bf 466} 3472
\bibitem{gil12}
Gilmore R, Somerville R, Primack J and Dominguez A 2012 {\it MNRAS} {\bf 422} 3189
\bibitem{kor18b}
Korochkin A and Rubtsov G 2018 {\it MNRAS} {\bf 481} 557
\bibitem{aha94}
Aharonian F, Coppi P and Voelk H 1994 {\it ApJ} {\bf 423} L5
\bibitem{dur13}
Durrer R and Neronov A 2013 {\it A\&A Rev.} {\bf 21}, 62
\bibitem{han17}
Han J L 2017 {\it Annual Review of Astronomy and Astrophysics} {\bf 55} 111
\bibitem{hac18}
Hackstein S et al. 2018 {\it MNRAS} {\bf 475} 2519
\bibitem{ber13}
Bernard D 2013 {\it Nucl. Instrum. Methods Phys. Res. Sec. A} {\bf 701} 225 [Erratum: Bernard D 2013 {\it Nucl. Instrum. Methods Phys. Res. Sec. A} {\bf 713} 76]
\bibitem{fun13}
Funk S and Hinton J A 2013 {\it Astropart. Phys.} {\bf 43} 348
\bibitem{cta13}
Bernlohr K et al. 2013 {\it Astropart. Phys.} {\bf 43} 171
\bibitem{act11}
Actis M et al. 2011 {\it Experimental Astronomy} {\bf 32} 193
\bibitem{ach13}
Acharya B et al. 2013 {\it Astropart. Phys.} {\bf 43} 3
\bibitem{lha14}
Cui S et al. (on behalf of the LHAASO Collaboration) 2014 {\it Astropart. Phys.} {\bf 54} 86
\end{thebibliography}
\end{document}